# Human-In-the-Loop for Bayesian Autonomous Materials Phase Mapping

*Felix Adams[1] (ORCID: 0000-0001-5802-1072)*
*Austin McDannald[2] (ORCID: 0000-0002-3767-926X)*
*Ichiro Takeuchi[1] (ORCID: 0000-0003-2625-0553)*
*A. Gilad Kusne[1,2] (ORCID: 0000-0001-8904-2087)*

1. Materials Science & Engineering Dept, University of Maryland, College Park MD
2. Materials Measurement Science Division, Material Measurement Laboratory, National Institute of Standards and Technology, Gaithersburg MD

**Abstract**
Autonomous experimentation (AE) combines machine learning and research hardware automation in a closed loop, guiding subsequent experiments toward user goals. As applied to materials research, AE can accelerate materials exploration, reducing time and cost compared to traditional Edisonian studies. Additionally, integrating knowledge from diverse sources including theory, simulations, literature, and domain experts can boost AE performance. Domain experts may provide unique knowledge addressing tasks that are difficult to automate. Here, we present a set of methods for integrating human input into an autonomous materials exploration campaign for composition-structure phase mapping. The methods are demonstrated on x-ray diffraction data collected from a thin film ternary combinatorial library. At any point during the campaign, the user can choose to provide input by indicating regions-of-interest, likely phase regions, and likely phase boundaries based on their prior knowledge (e.g., knowledge of the phase map of a similar material system), along with quantifying their certainty. The human input is integrated by defining a set of probabilistic priors over the phase map. Algorithm output is a probabilistic distribution over potential phase maps, given the data, model, and human input. We demonstrate a significant improvement in phase mapping performance given appropriate human input.

**Introduction**
New materials with improved properties are needed to solve modern engineering challenges, but with each year, discovering such materials becomes more difficult (*1*). As researchers exhaust simple-to-make materials, they are driven to investigate materials systems with ever more complex synthesis and processing conditions. However, with each new synthesis or processing parameter, the number of potential materials (and materials experiments) grows exponentially. The Edisonian trial-and-error approach cannot scale appropriately (*2*). Several strategies for accelerating materials discovery exist. High-throughput combinatorial synthesis creates hundreds of diverse material samples at once, providing a geometric acceleration. High-throughput computational models also provide a geometric speed-up. Both these approaches cannot scale to address the exponentially growing search space.

Machine learning (ML) offers a solution with the promise of 1) extrapolating knowledge from past experiments (both in lab and in silico) throughout the vast search space and 2) using the extrapolation results to identify the set of subsequent experiments that may accelerate the user toward their goals. Experiment selection is performed using active learning – the machine learning field of optimal experiment design (*3*). By focusing on information-rich experiments, an exponential acceleration in research is achievable (*2*). Autonomous experimentation (AE) combines such machine learning algorithms with automated research equipment to perform experiment design, execution, and analysis in a closed loop.

The successful implementations of AE span diverse areas of materials exploration and optimization. For example, AE has accelerated process optimization for chemical vapor deposition of carbon nanotubes using in-situ Raman data (*4*) and chemical mixture optimization for improved optical thin films (*5*). AE has accelerated the process of physics model and parameter determination through control of neutron scattering measurement experiments of magnetic materials (*6*). AE was also demonstrated to accelerate materials discovery, guiding x-ray diffraction measurements to learn composition-structure-property relationship of phase change memory materials and identifying a novel material. This AE study has resulted in the first (and to the best of the knowledge of the authors, the only) discovery of a novel best-in-class material, specifically a phase change memory material (*7*).

Some of these and other past AE systems combine knowledge from physics theory, computation, and experiments. Prior physics theory was integrated into ML algorithms through diverse methods including constraint programming, probabilistic programming, and probabilistic priors. For example, constraint programming was used to integrate Gibbs phase rule into AE x-ray diffraction-based phase mapping systems (*7*, *8*). Probabilistic programming was used to integrate magnetic theory and neutron scattering theory into an AE neutron scattering-based system for Neel temperature determination of an antiferromagnetic material (*6*). Additionally, Bayesian analysis probabilistic priors have been used to integrate knowledge from previous density functional theory calculations and experiments into a probabilistic model controlling AE phase mapping (*7*).

However, AE systems have yet to systematically include human expertise into the AE closed-loop (*2*). While algorithms may outperform humans at fast, detailed, high-dimensional, tedious tasks, human experts win out at hard-to-define tasks that require insight beyond the scope of the given task. By combining human expertise with AE ML, one can potentially overcome the individual weaknesses of man and machine to boost AE performance. Such human-computer interaction falls under 'human-in-the-loop', a major subfield of computer science (*9*).

There is an added benefit to overcoming the challenge of human knowledge integration into the AE loop. To ensure success of the AE and human expert collaboration, the expert must trust that there is value in the interaction. The expert must also understand the algorithm's current state to know what input to provide. These require tackling the ever-present challenges of machine learning trustworthiness and interpretability [5]. Ensuring trustworthiness and interpretability is essential to future adoption of such systems in industry research pipelines.

Previous work utilizing human input in AE and materials informatics focused on the human as a client, with three types of interaction: 1) ML-based experiment recommendations, 2) ML-directed human computation, and 3) human-based algorithm parameter updates. ML-based recommendation engines have guided research into diverse materials including phase change memory materials (*10*), high entropy alloys (*11*), and photovoltaics (*12*). Here ML selects subsequent experiments, the researcher performs the experiments, and the researcher then updates the ML's database with the collected data. Regarding ML-guided human computation, distributed human computing was used to identify x-ray diffraction peaks associated with single phases (*13*). This work used a graphical user interface to display x-ray diffraction data to human clients who then identified the location of peaks. An ML algorithm has also directed human-run optical bandgap calculations for ellipsometry measurements (*7*). Toward human-based algorithm parameter updates, human input was used by the first iteration of AE-based carbon nanotube process optimization to regularly update experiment design goals to avoid local optima (*4*).

For this work we focus on collaborative interaction between human and ML forming a human-AI team, where both provide knowledge of a similar type. In particular, the ML and human expert combine their predictions (i.e., expectations) of future experiment outputs and the utility – i.e., quantified value, they assign to these experiments. The application is AE accelerated phase mapping – guiding subsequent experiments to maximize knowledge of a composition-structure phase map through optimal selection of materials compositions to be characterized by x-ray diffraction. This challenge was selected as boosting the performance of AE phase mapping has been shown to accelerate materials discovery (*7*, *14*).

We integrate human input through a set of probabilistic priors as well as additive and multiplicative functions to combine ML and human knowledge of 1) phase map prediction, i.e., structure data extrapolation (rather than phase analysis of x-ray diffraction data), and 2) subsequent experiment utility for experiment design. The human expert interacts with the algorithm through a graphical user interface that visualizes ML phase map predictions and the utility of choosing any experiment as the subsequent experiment to perform experiment as defined by the acquisition function. The work here is demonstrated through a closed-loop AE simulation. We first present the phase mapping model, then discuss the methodology for human expertise integration, and finally present a demonstration and efficacy quantification of human input on autonomous phase mapping.

**Methods**

We present a graphical user interface where users provide input to the AE phase mapping algorithm toward two tasks: 1) improving phase map predictions, and 2) improving selection of subsequent experiments, i.e., experiment design. For the first task, at the current AE iteration, the algorithm presents the user with the most likely phase map given the data, and the user provides their input by drawing boundaries and regions on this phase map. Toward the second task, the algorithm presents the user with its utility for potential subsequent experiments, and the user draws a boundary around their own region of interest. We describe the phase mapping algorithm and then discuss the math behind the human input integration.

*Autonomous Phase Mapping Algorithm*

The phase mapping algorithm is based on the AE 'closed-loop autonomous exploration and optimization' (CAMEO) algorithm (*7*), which guides experiments to learn the composition-structure-property relationship and then exploits that knowledge to hone in on user-goal-based optimal materials. The implementation used here focuses on learning the composition-structure property relationship, i.e., the composition phase map. Here $x_i = [x_{i,0}, x_{i,1}, ..., x_{i,N}]$ is a vector describing the independent variable, in our case, the material synthesis and/or processing conditions, e.g., the composition, and $y_i$ describes the x-ray diffraction data associated with $x_i$. The steps of the algorithm are given by:
1) *Analyze past data.* Estimate a phase map for previously characterized materials. Spectral clustering is used to estimate materials likely to belong to the same phase region. The similarity $w_{i,j} = \exp[-d_{i,j}^2/\sigma^2]$ is used, where $d_{i,j} = 1 - \frac{\text{dot}(y_i, y_j)}{\text{norm}(y_i) * \text{norm}(y_j)}$ is the cosine metric between $y$ the x-ray diffraction patterns (in the form of peak intensity vs. 2θ) for materials with compositions $x_i$ and $x_j$ and $\sigma = 1$. The number of expected phase regions is set to 5.
2) *Extrapolation.* Extrapolate phase region knowledge from x-ray diffraction characterized materials to yet-to-be measured materials. Here, Gaussian process classification (GPC) is used to extend knowledge of phase region labels from the analysis step across the target composition space. A 2-dimensional Matérn 3/2 kernel is used, with a fixed variance of 1 and fixed length scale of 0.2 elemental fraction.
3) *Optimal experiment design.* Subsequent experiments target materials that maximize knowledge of the phase map. The utility of each experiment toward this goal is computed with the acquisition function $\alpha$. For this demonstration, experiments are selected that have maximum entropy in $p_k$ the probability distribution across potential phase region labels:

$$\alpha(x) = \sum_k p_k(x) \ln p_k(x), \quad Eq\ (1)$$

Larger entropy indicates greater uncertainty in phase region label and thus correlates with the position of phase boundaries.
4) *Closing the loop.* Here an AE closed-loop system is simulated by identifying the next material to measure from step 3 and then calling the associated x-ray diffraction pattern from a previously populated database.

*Human Knowledge-based Probabilistic Prior via Gaussian Process Classification*

Gaussian process classification (GPC) is a probabilistic classification method: rather than just outputting a classification estimate, it provides a full posterior probability including a prediction of the most likely classification and a quantification of uncertainty. Underlying GPC is the search for a set of latent functions $f_i$, each used to quantify the probability of input $x_j$ belonging to class $i$ vs any other class. The latent functions are sampled from a multivariate normal distribution $f_i \sim N(\mu_i, K)$ where $\mu_i$ is the distribution mean and $K$ is the covariance. Together, $\mu_i$ and $K$ define a prior over the form of $f$. Typically, $\mu_i = 0$, due to a lack of prior knowledge of the mean behavior of $f_i$. $K$ is designed to impose probabilistic prior assumptions of $f_i$'s covariance behavior. For instance, if $K$ is selected to be a radial basis function (RBF), $f_i$ must be a smooth function, with a larger kernel length scale resulting in a larger contiguous regions of similar class. One can thus design the mean and kernel for off-the-shelf GPC algorithms to integrate human knowledge into the probabilistic prior.

Kernels are typically applied to the input vector data $x$, i.e., $K = k(x_i, x_j)$. Thus, one must first encode the human input as a vector and introduce it through $x$. Assuming vectorized human input $v_{x_i} = [v_{i,0}, v_{i,1}, ..., v_{i,M}]$ one can then modify $x_i$, giving $x'_i = [x_{i,0}, x_{i,1}, ..., x_{i,N}, v_{i,0}, v_{i,1}, ..., v_{i,M}]$. The common kernel $K = k(x_i, x_j)$ is updated to $K' = k(x_i, x_j, v_i, v_j)$, giving a kernel that integrates human input as a probablistic prior. For example, one can employ the additive kernel $K = \beta_x g_x(x_i, x_j) + \beta_v g_v(v_i, v_j)$ or the multiplicative kernel $K = g_x(x_i, x_j) g_v(v_i, v_j)$. The additive kernel mirrors a logical OR statement, where the kernel has a large value if either $g_x$ or $g_v$ have a large value. Additionally, if $\beta_v \gg \beta_x$, the human input prior dominates the prediction. $\beta_v/\beta_x$ can thus be selected to quantify human certainty in their input. The multiplicative kernel mirrors a logical AND statement, where the kernel has a large value if both $g_x$ and $g_v$ have large values and vice versa.

*Human-in-the-Loop: User Input Integration Methods*

We created a graphical user interface which allows a user to provide their input toward three topics: 1) cohesive phase segments, i.e., regions of the synthesis and processing space with materials likely to belong to the same phase region, 2) likely phase boundaries, and 3) regions-of-interest - portions of the search space that may contain pertinent information. At each iteration, the predicted phase map - the GPC extrapolation estimate, is presented to the user. If the user indicates they would like to provide input by pressing a button, the current iteration finishes, the latest data and predictions are displayed, and the user provides input through the interface. This method assumes the user is presented with a set of unmeasured sample materials. This set may be chosen as a grid over the composition space or drawn from a set of material samples previously define (i.e., pool sampling (*3*)). For this work, the set of all materials are drawn from a set of previously defined materials.

1. **Cohesive Phase Segment**

Here, the user draws a boundary around a segment of the composition space they believe contains materials sharing the same phase region. Materials with composition found inside the bounded segment are given $v_i = V_{in}$ and materials with composition outside the bounded segment are described by $v_i = V_{out}$, where $V_{in} \neq V_{out}$. $g_v$ is then given by:

$$g_v(v_i, v_j) = \begin{cases} 1 \text{ if } v_i = v_j \\ 0 \text{ if } v_i \neq v_j \end{cases}, \quad Eq\ (2)$$

i.e., $g_v = 1$ if the composition of two materials both fall either inside the segment or outside the segment, and a value of 0 otherwise. The GPC kernel is given by:

$$K = \beta_x RBF(x_i, x_j) + \beta_v g_v(v_i, v_j), \quad Eq\ (3)$$

For the examples provided $\beta_v = \beta_x$, indicating a user with 50 % confidence. They are just as confident in their input as the input provided by the ML algorithm.

2. *Phase Boundary*

Materials on one side of the boundary or the other are treated in the same way as materials inside/outside the drawn phase region in the previous case, with the same kernel applied.

   3. *Region of Interest*

The user draws a boundary around a portion of the composition space believed to contain materials of high interest. The boundary defines a region of increased interest with

$$r(x) = \begin{cases} 1 & if\ \boldsymbol{x}_i \text{ is inside boundary} \\ 0 & if\ \boldsymbol{x}_i \text{ is outside boundary} \end{cases}, \quad Eq\ (4)$$

The learned acquisition function $\alpha$ is then modified by either $\alpha' = \alpha * r(\boldsymbol{x})$ or $\alpha' = \beta_\alpha \alpha + \beta_r * r(\boldsymbol{x})$, where $\beta_\alpha$ and $\beta_r$ are constants. The former model restricts the subsequent search to within the boundary. The latter model boosts utility values within the region. The selection of $\beta$ coefficients can be used to normalize the two contributions, ensuring a balanced impact of model output (encoded in $\alpha$) and user input. Here we demonstrate the former.

*Algorithmic Process*

Figure 1 demonstrates the algorithmic process for human input integration. Each iteration includes a computational analysis of previously collected data and then the measurement of subsequently selected materials. For this example, the autonomous phase mapping algorithm is allowed to select a sequence of four materials to investigate. Analyses of the collected data is completed and displayed (the left plot) and the fifth material to investigate is selected. At this point the user indicates their desire to provide input by pressing a button, pausing the algorithm. The user indicates a cohesive phase segment by drawing a quadrilateral (the central plot). The user then indicates that they have completed with their input, restarting the algorithm. Their input thus impacts analysis in the next iteration (the right plot) and the resulting selection of the sixth material to investigate.

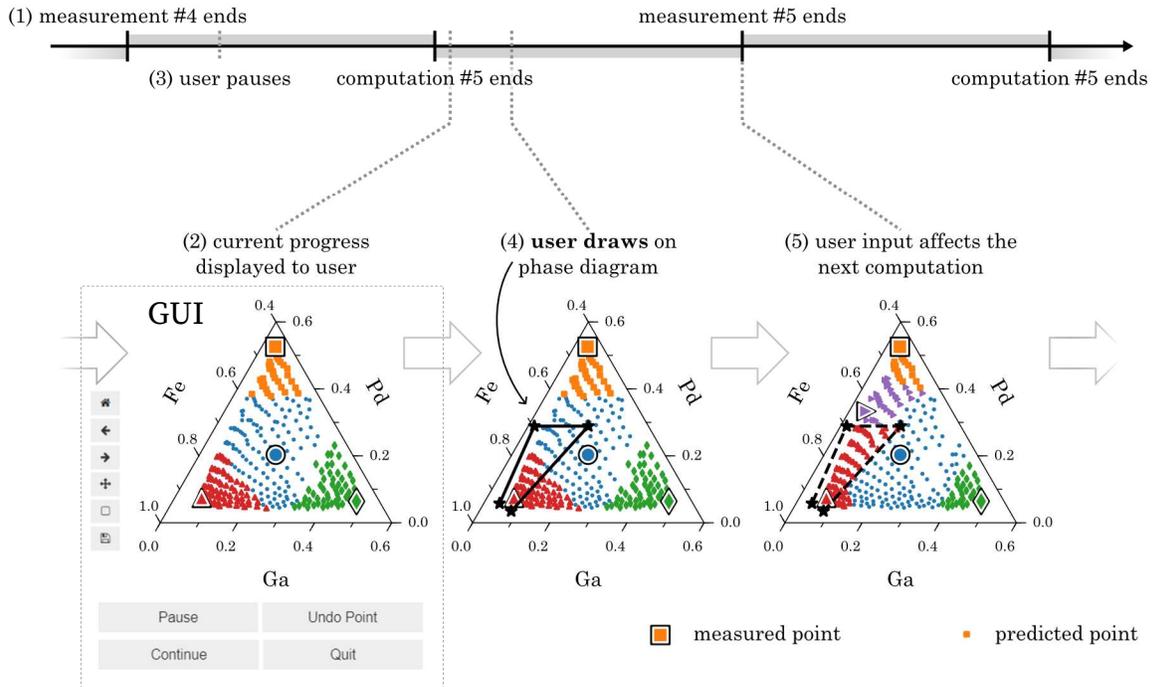

**Figure 1.** Example illustration of how user input is captured and used in the autonomous algorithm. The ternary plots show the grouping (i.e., clustering) of materials into potential phase regions within a ternary system A-B-C. Markers indicate individual materials. Materials with markers of the same color and shape indicate that they are assigned to the same phase region. 1) In this example, the machine learning algorithm selects four materials to investigate, 2) analyzes the collected data, displaying the results in the graphical user interface (GUI, left plot), and the fifth material to investigate is selected. 3) The user indicates their desire to provide input, pausing the algorithm. 4) The user indicates a cohesive phase segment by drawing a quadrilateral (the central plot) and then indicates that they have completed their input, restarting the algorithm. 5) The user input impacts analysis in the next iteration (the right plot) and the resulting selection of the sixth material to investigate. Analysis prediction is indicated by a color-coded solid marker and measured materials are indicated with an outline marker. Phase region estimate is indicated by solid markers of different color and shape.

## Results and Discussion

To determine how human input affects the autonomous experiment, we performed simulated autonomous experiments using the Fe-Ga-Pd ternary diffraction dataset (*15*). We compared the model's output in 4 cases: 1) a control with no human input, 2) a boundary input, 3) a cohesive segment input, 4) and a region-of-interest input. For human inputs, we use a phase boundary and a phase region (cohesive phase segment) similar to (though not exactly the same as) those present in the true phase map as provided by References (*15*, *16*), using the same $2\theta$ range. In Figure 2 we show the algorithm's phase diagram predictions and acquisition function values for the four described cases. We analyzed the effect of human input early in the autonomous experiment (only five materials have been characterized) so that the model's predicted phase map is rudimentary and does not match the ground truth.

The left column of Figure 2 shows the model's predicted phase diagram after five compositions have been measured for each of the four cases. As prediction is Bayesian, these are the most likely phase diagrams given the model, data, and any human input. For the control case, the algorithm has no human input and no information about the material system between the measured materials, so the phase boundaries are estimated as half-way between measured materials. For the boundary and region input cases, when the user provides information about the phase structure between measured points, the phase boundaries align with the provided boundary or region. Note that the boundary input changes the predicted phase region of fewer materials than the region input.

Human input for boundary and region can have varying impact on the acquisition function, where the material with the maximum acquisition value is the material with highest utility. The top right plot of Figure 2 shows a heatmap of the entropy acquisition function calculated for the next iteration of the control. The center right and bottom right plots of Figure 2 show the change in the acquisition function values due to human input. The boundary input has a smaller effect on the acquisition function values than the region input corresponding to the smaller change in the phase diagram as mentioned above. The region input locally decreases phase region uncertainty, decreasing the likelihood that these materials are chosen for subsequent characterization. For human input of their region-of-interest, human input only impacts the acquisition function, reducing the utility of materials outside the region of interest while maintaining the relative utility of materials within the region.

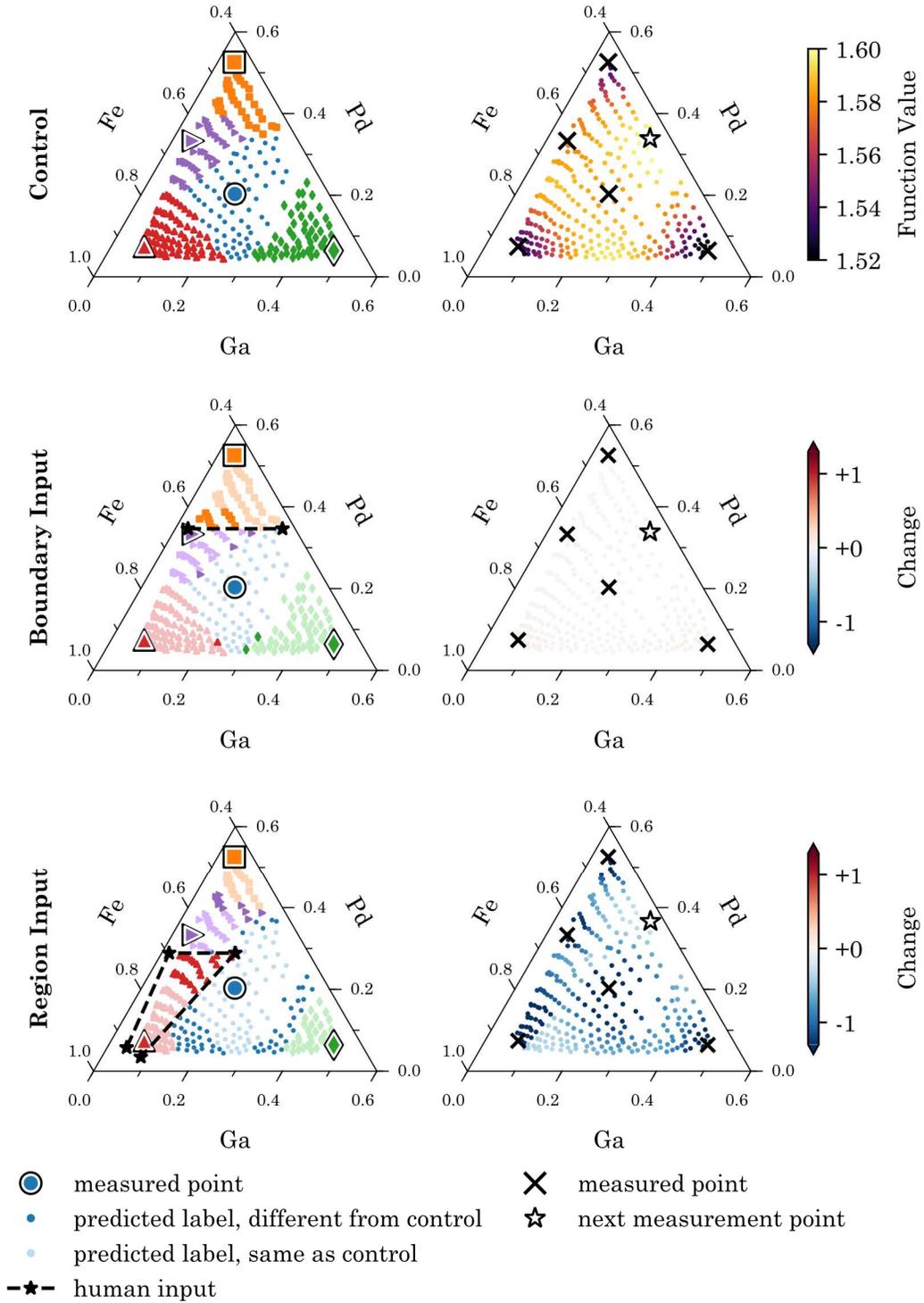

**Figure 2.** Example clustering and acquisition function results after four cases of human input: 1) control - no user input (first row), 2) a phase boundary input (second row), and 3) a phase region input (third row). The first column shows the human input label with the resulting predictions. The second column shows the result on the acquisition function. Here materials indicated by the same markers (color, shape) indicate materials assigned to the same phase region. Outline markers in the left column and 'x' markers in the right column indicate measured materials. Stars ⋆ and dashed lines indicate points and boundaries or regions defined by the user.

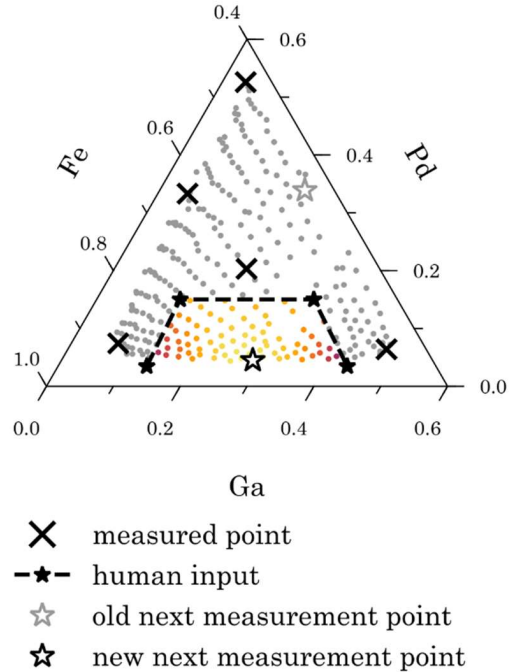

**Figure 3.** The effect of the third type of human input: the region of interest. Compared to the top right plot of Figure 2, all of the acquisition function values outside of the user – defined region are optionally reduced or neglected. In the latter case, the next measurement point is chosen from inside the region.

      To quantify the effect of the human input, we performed 500 trials for each of the cases of boundary and region input. For each trial, 5 initial materials are randomly selected, the algorithm analyzes the data for these materials, human input is provided identical to that shown in Figure 2 (boundary and region inputs, respectively), and the resulting phase map estimate is compared to the ground truth using the Fowlkes-Mallows index (*17*). A two-sided Mann - Whitney U test is used to determine if the difference in the Fowlkes - Mallows index due to human input is statistically significant (*18*). The Mann - Whitney U test statistically evaluates whether two distributions are different by comparing observation probabilities between two sample populations. Here the Mann - Whitney U tests indicate that both inputs provide a significant ($p < 0.05$) effect on the Fowlkes - Mallows index. The data from the random trials are summarized in Table 1 below.

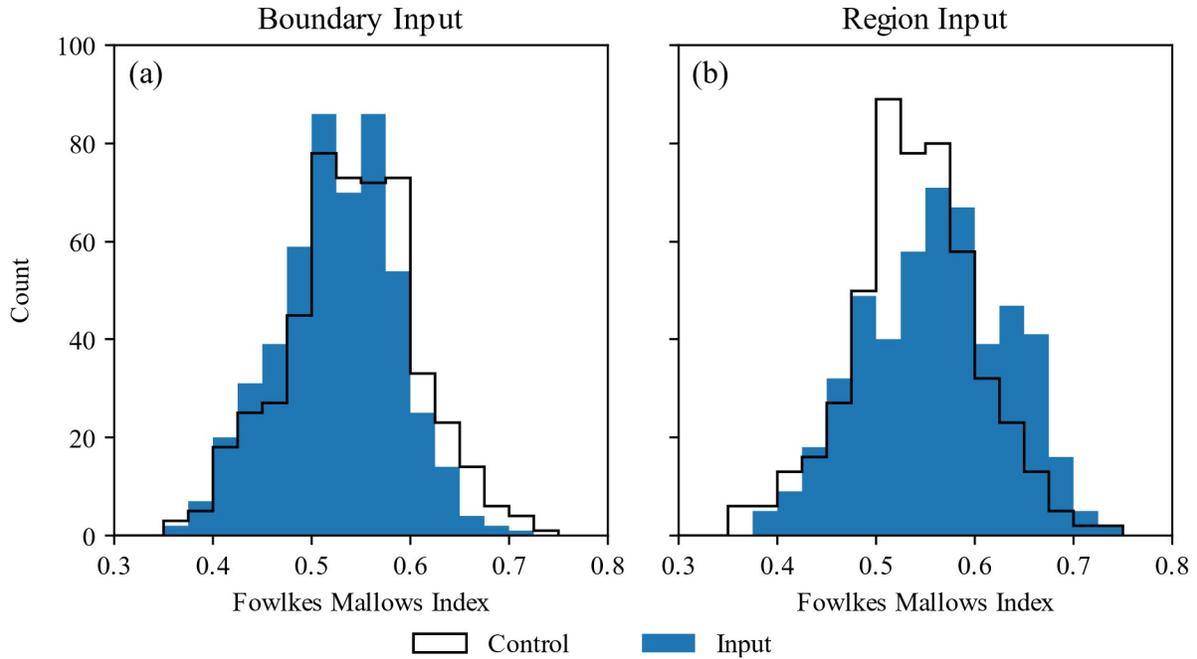

**Figure 4.** Histogram of the Fowlkes Mallows Index values used to perform the Mann – Whitney U test for two human input cases: boundary input (a) and region input (b). 500 trials are run for each case with each trial beginning with the random selection of 5 materials.

|  | Sample Median FMI | Trials | U value | $p$ value |
| --- | --- | --- | --- | --- |
| Control (No Input) | 0.54 | 500 | – | – |
| Boundary Input | 0.53 | 500 | 107,923.5 | $1.8 \times 10^{-4}$ |
| Region Input | 0.56 | 500 | 148,534 | $2.6 \times 10^{-7}$ |

**Table 1.** Results of the Mann - Whitney U test performed on the 500 trials for both boundary and region inputs and the resulting Fowlkes Mallows indices.

    The effect of the user input depends on when the input is provided. Figure 2 shows the results when only 5 materials have been characterized, which is relatively early in the phase-mapping process. The user input has a larger impact on the predicted phase map as the model has less information to make its prediction. The impact decreases with more structure data. For these experiments, user input does not affect the phase labels of the measured materials, as user input is not integrated into the AE analysis step.

    Th same described method of integrating user input through the algorithm kernel can be applied to the analysis step. This experiment is performed with spectral clustering, allowing user input to split and combine phase regions. Here, the kernel calculated from the user input is added to the cosine measure-based similarity matrix. Phase boundary or region

input can thus result in spectral clustering assigning different phase region labels to materials on either side of the user-provided boundary or inside and outside a region. Figure 3 shows the effect of the phase boundary and phase region inputs on the results of the spectral clustering with 20 measured compositions. Note that the specific label for each phase region (orange square, blue circle, etc.) is randomly assigned.

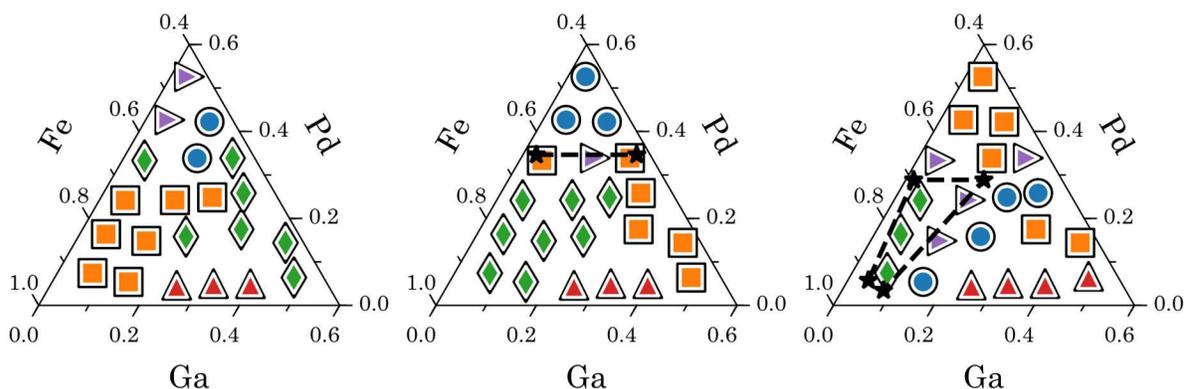

**Figure 5.** The effect of adding the user input kernel to the spectral clustering similarity matrix.

## Conclusions

We have described a graphical user interface that presents a user with interpretable figures for ML analysis and predictions and allows the user to impart their knowledge to the algorithm. Through the act of drawing boundaries and regions on ML-computed phase maps, the user significantly impacts data analysis, algorithm prediction, and optimal experiment design. Such interpretable human input methods are essential for the future of autonomous systems in the lab, as they allow experts to work in human-ML teams and establish trust.

## Data and Code
Link to Github repository: https://github.com/fadams-umd/CAMEO-HITL